\documentstyle[aps,prb,twocolumn,overcite,epsfig]{revtex}
\def\be{\begin{equation}}
\def\ee{\end{equation}}
\def\eps{\epsilon}
\def\cl{c_{\rm light}}
\def\epseff{\epsilon}
\def\mueff{\mu}
\def\neff{n}

\begin{document}
\title{Resonant and anti-resonant frequency dependence of the 
effective parameters of metamaterials}

\author{T. Koschny$^1$, P. Marko\v{s}$^{2,3,*}$, D. R. Smith$^4$ and C. 
M. Soukoulis$^{1,3}$}

\address{%
$^1$ Research Center of Crete, 71110 Heraklion, Crete, Greece\\
$^2$ Institute of Physics, Slovak Academy of Sciences, 845 11 
Bratislava, Slovakia\\
$^3$ Ames Laboratory and Dept. Phys. and Astronomy, Iowa State 
University, Ames, Iowa 50011\\
$^4$ Dept. Phys., University of California, San Diego, 9500 Gilman 
Drive, La Jolla, CA 92093-0319
}

\maketitle

\abstract{%
We present a numerical study of the
electromagnetic response of the  metamaterial elements that are used
to construct materials with negative refractive index.
For an array of split ring resonators (SRR) we find
that the resonant behavior of the effective magnetic permeability
is  accompanied by an  anti-resonant behavior of the
effective permittivity.  In addition, the  imaginary parts of the effective
permittivity and permeability are opposite in sign.
We also observe an identical resonant
versus  anti-resonant frequency dependence
of the effective materials parameters  for
a  periodic array of thin metallic wires with cuts
placed periodically along the length of the wire, with roles of the 
permittivity
and permeability reversed from the SRR case.
We  show in a simple manner that the finite unit cell size is
responsible  for the anti-resonant  behavior.
}

\medskip


\medskip

The recent development of metamaterials with negative refractive index -
or double-negative (DNG) metamaterials
\cite{ZH}
has   confirmed that
structures can be fabricated that can be interpreted
as having both  a negative effective permittivity $\epseff$
and a negative effective permeability $\mueff$ simultaneously.
Since the original microwave  experiment of Smith {\it et al.}, \cite{Smith-1}
various new samples were
prepared \cite{Smith-2,Li}, all of which have been shown to exhibit a pass band
in which  the permittivity and permeability are both negative.
These materials have been used to demonstrate negative refraction of 
electromagnetic waves,
\cite{Shelby}
a phenomenon predicted by Veselago
\cite{Veselago}. Subsequent experiments \cite{Claudio-2}
have reaffirmed the property of negative refraction,
giving strong support
to the interpretation that these metamaterials can be correctly described
by negative permittivity and negative permeability.
\cite{Pendry-1,Pendry-2}

There is also an increasing amount  of numerical work  \cite{MS-2,numerics,ms}
in which the  transmission and reflection of electromagnetic wave
is calculated for a finite length of metamaterial.
For a finite slab of continuous material, the complex transmission 
and reflection
coefficients are directly related to the refractive index $n$ and impedance $z$
associated with the slab, which can in turn be expressed in terms of 
permittivity
$\epsilon$ and permeability $\mu$. A retrieval procedure can then be 
applied to find material parameters for a finite length of 
metamaterial, with the assumption that the material can be treated as 
continuous.
A retrieval process
was applied  in Ref. \cite{SSMS}, and  confirmed
that a medium composed of split ring resonators (SRRs)  and wires could indeed
be characterized
by effective $\epseff$ and $\mueff$ whose real parts were both negative over
a finite frequency band, as was the real part of the refractive index $\neff$.

The retrieval process, however, uncovers some
unexpected effects. For the SRR medium, for instance,
the real part of the effective permittivity
$\epseff'$
exhibits an  anti-resonant
frequency dependence in the same frequency region where the permeability
undergoes its resonance.
This anti-resonance can be seen in the composite SRR+wire negative 
index medium as well.
The anti-resonance in the real part of the permittivity is also 
accompanied by anti-resonant behavior in the imaginary part of the 
permittivity $\epseff''$, which exhibits an absorption peak opposite 
in sign to that of the imaginary part of the permeability.
Assuming that waves have a time dependence of $\exp(-{\rm i}\omega 
t)$, one would expect the imaginary parts of both the permittivity 
and permeability to have positive  values at
all frequencies, since the material is passive. \cite{PE,Landau}

The aim of present paper is to show that the  anti-resonant behavior 
of the material
parameter  is an intrinsic  property of a metamaterial, a consequence 
of  the finite
spatial periodicity. To illustrate this point,  we present
the retrieved material parameters for two types of metamaterial media 
that have been used to form negative refractive index composites:
the first medium comprises an array of SRRs that exhibit
a resonant  permeability and anti-resonant permittivity. The second
medium comprises an   array of
cut metallic wires,  which exhibits resonant permittivity and 
anti-resonant permeability.

\smallskip

\begin{figure}
\begin{center}
\epsfig{file=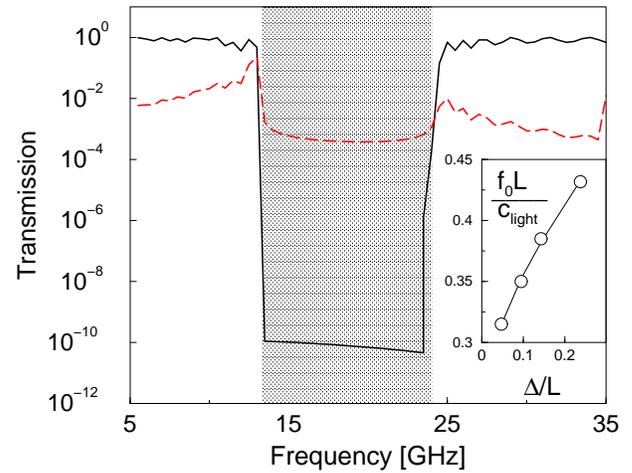,width=0.45\textwidth}
\end{center}
\caption{Transmission of the EM wave through the periodic lattice of 
thin metallic wires.
Wires are parallel with the $y$ axis, and the EM wave, polarized with
$\vec{E}\parallel y$, propagates in the $z$ direction.
The system is infinite in the $x$ and $z$ direction, and
60 rows of wires are considered along the propagation direction.
The structure  is characterized by four length parameters:
the wire thickness $w$,   gap $\Delta$,
the wire length $L$ and the lattice period (mutual distance of wires) 
$a$. In the present
simulations, the lattice constant
is $a=3.66$ mm, the thickness of the wire is $w=0.33$ mm, the length 
of the wire is
$L=7$  mm  and
the cut of the wire is $\Delta=0.33$ mm.
Data show the drop of the transmission (solid line)
at frequency $f_0\approx 13.35$ GHz.
Dashed line is absorption.
We found that the resonant frequency
$f_0$ is almost independent  on the lattice constant $a$
($1.66\le a\le 5$ mm).
However,  it depends strongly on the  gap  $\Delta$. In
inset shows how the frequency $f_0$
depends on the ratio  $\Delta/L$.
Solid line is the fit 
$f_0=\frac{\cl}{\sqrt{2\pi}L}\left[a_0\ln(L/\Delta)+a_1\right]^{-1/2}$
with $a_0=0.48$ and $a_1=0.17$.
Plasma frequency $f_p$ depends on the lattice period $a$.
For $a=3.66$ mm we found numerically $f_p\approx 24.5$ GHz
which agrees  with predictions of Sarychev and Shalaev 
\cite{Sarychev}, indicating that the value of
the plasma frequency is not influenced
by the wire cut.
}
\label{Btw}
\end{figure}

We used the transfer matrix method to simulate numerically the
transmission of the electro magnetic waves through the metamaterials.
Transmission data for an array of SRR were published elsewhere 
\cite{MS-2,ms} and will not be repeated
here.
To the best of our knowledge, the  analysis of the array of cut wires 
has not been presented
in the literature yet.
Fig. \ref{Btw} shows the frequency dependence of the transmission and 
absorption for an array of
cut wires.  As expected,
the  system exhibits a  band gap  for frequencies $f_{0}<f<f_p$.
Here, $f_0$ is the resonance frequency and $f_p$ is the plasma frequency.
The transmission data are similar to that for an array of the SRR. However,
  contrary to an
array of SRR, the system exhibits the resonant behavior of the 
effective permittivity at
$f=f_0$.

\medskip

 From the transmission and reflection data we calculate the effective 
permittivity and permeability.
Details of the method were published elsewhere \cite{SSMS}.
The method is based on the assumption that the system is homogeneous. 
Textbook formulas for the
transmission and reflection of the slab of  width $d$ are then 
inverted to obtain
effective impedance $z$ and effective refractive index $n$.
Permittivity and permeability are obtained from relations
\be\label{one}
n=\sqrt{\eps\mu}~~~{\rm and}~~~z=\sqrt{\mu/\eps}.
\ee

\medskip

\begin{figure}
\begin{center}
\epsfig{file=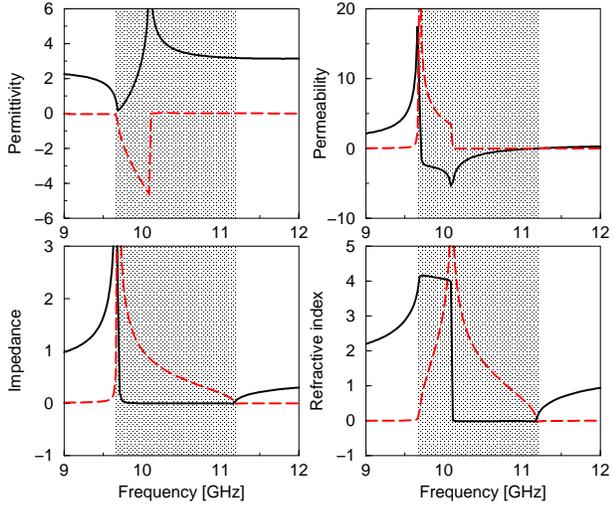,width=0.45\textwidth}
\end{center}
\caption{Effective parameters for an array of split ring resonators 
(solid lines: real part,
dashed lines: imaginary part).
The resonant behavior of the effective permeability $\mueff$ at 
frequency $f_0=9.66$ GHz is clearly visible.
Shaded area shows the resonance frequency interval in which
$\mueff'$ is negative.
Note the  anti-resonant behavior of $\epseff$. Note also that 
$\epseff''$ is negative.
The sharp discontinuity in $\epseff''$ is due to the extremely fast 
decrease of $\neff'$
to zero in the right part of the resonance interval.
The size of SRR is 3 mm, the size of unit cell is $L_x\times 
L_y\times L_z=3.33\times 3.66\times 3.66$
mm. The SRR lies in the $x=0$ plane, EM wave propagates along the $z$ 
direction  and is $E\parallel y$
polarized.
}
\label{0yL}
\end{figure}

\begin{figure}
\begin{center}
\epsfig{file=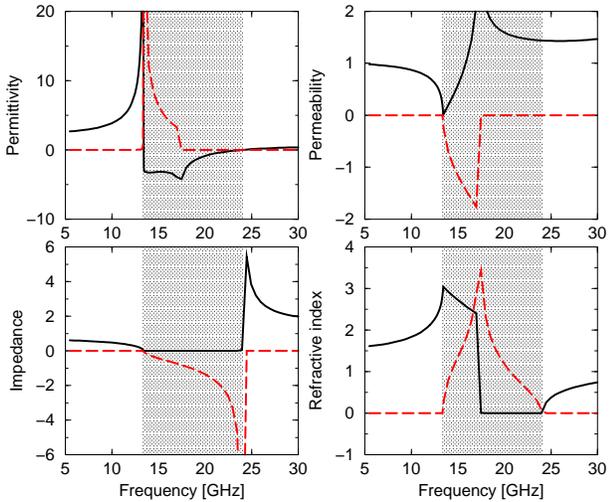,width=0.45\textwidth}
\end{center}
\caption{Effective parameters of the periodic lattice of cut wires.
Resonant behavior of the effective permittivity $\epseff$ as well as 
anti-resonant
behavior of effective permeability $\mueff$ are clearly visible.
The real part
of permeability $\mueff'$  is zero at $f_{0}$,  and
the imaginary
part of the permeability $\mueff''$ is negative for $f>f_{0}$.
}
\label{Btw-11f}
\end{figure}

Typical frequency dependence of the effective parameters of an
array of SRR and array of cut wires  are  shown in fig. \ref{0yL} and 
\ref{Btw-11f}, respectively.
Both structures exhibit qualitatively the same behavior: there is a 
resonant frequency interval,
in which one effective parameter is negative ($\mueff'$ for SRR, 
$\epseff'$ for cut wires).
The resonant behavior of this parameter at the
left border of the band gap is clearly visible. The
frequency dependence of the  second parameter is anti-resonant. Its real part
decreases to zero at $f=f_0$ and  imaginary part  is negative for $f>f_0$.
Notice the qualitative similarity of both systems:
data are qualitatively the same, and differ only
in the exchange of $\epseff$ and $\mueff$ (which is equivalent to the 
transformation $z\to 1/z$).

\medskip

The anti-resonant behavior of the effective parameter has its origin in the
finite lattice period $a$ associated  with
the metamaterial structure.
One manifestation of the lattice periodicity is that there is a 
maximal wave number,
given by $k_{\rm max}=\pi/a$. \cite{footnote1}
If we assume that the metamaterial can be treated as a continuous medium
  with an index of refraction $n$, then the definition of $n=\cl k/\omega$
shows that $n$ is necessarily bounded.
The generic $\omega(k)$ dispersion diagram of a resonant periodic structure
results from the coupling of a dispersionless resonance curve at the 
resonant frequency
$\omega_0=2\pi f_0$ and the light line $f=\cl/k$ (see Ref. 
\cite{Smith-1} for details).
The result is a lower branch that extends from zero frequency to the resonance
frequency $f_0$, followed by a band gap that extends
from $f_0$ to $f_p$, followed by an upper branch that extends upwards 
in the frequency from
$f_p$.

Because the resonant frequency of a typical metamaterial element
implies a free space wavelength much longer than the unit cell size, 
an effective medium
approach has been applied that results in a characterization of metamaterial
in terms of bulk $\epsilon$ and $\mu$.
The square root of the product $\eps\mu$ is the refractive index, which must be
consistent with that determined from the dispersion diagram.
To understand the generic properties associated with a resonant 
system in a periodic structure, we start by considering the 
dispersion characteristics of the system near the resonant
frequency, where we have
\be\label{oo}
\omega\sim\omega_0-\frac{1}{\alpha^2}\left(k-\frac{\pi}{a}\right)^2
\ee
where $\alpha$ is a real number. Solving Eq. \ref{oo}  for $k$ and 
using $n=\cl k/\omega$,
we find an approximate expression for the refractive index:
\be
n(f)\approx \frac{\cl}{2\pi 
f_0}\left(\frac{\pi}{a}-\alpha\sqrt{\omega_0-\omega}\right).
\ee
The maximum value of the refractive index at the resonance frequency is thus
\be\label{nmax}
n_{\rm max}=\frac{\cl}{\omega_0}\frac{\pi}{a}
\ee
determined only by the resonant frequency of the element and the periodicity.

For the composite structures analyzed in this paper,
Eq.  \ref{nmax} gives  $n_{\rm max}=4.24$
for an array of SRRs
and  $n_{\rm max}=3$
for the cut wires array.
When comparing $n_{\rm max}$ with data presented in figs. 2 and 3, we see that
$n'$ indeed does not exceed these limits.

Assume that the effective parameter $x$ ($x$ represents
the  effective permittivity for the cut wires system
and the  effective permeability for the  SRR system)
  exhibits a resonant form corresponding to
\be\label{mueff}
x(f)=1-\frac{Ff^2}{f^2-f_0^2+i\gamma f}.
\ee
where $\gamma$ is a damping factor, and $F$ is a filling
factor (fraction of volume of the metallic components).
At resonance, the imaginary part $x''(\omega)$ becomes
\be
x''(\omega)=(1-F)\frac{\omega_p^2-\omega_0^2}{\gamma\omega_0}>0
\ee
($\omega_p^2=\omega_0^2/(1-F)$)
and is greater than zero, as expected.

If we  require that the index $n$ calculated form the dispersion curves
be consistent with the bulk permittivity and permeability parameters, 
than we must have
$n(\omega)=\sqrt{x(\omega)y(\omega)}$. Near the resonance frequency 
this implies that
the second effective parameters
$y(\omega)$  behaves as
\be\label{yy}
y(\omega)=\frac{n^2(\omega)}{x(\omega)}.
\ee
Comparing the expression for  $y(\omega)$ with that for $x(\omega)$ 
we see that the poles
and zeros of $x$ and $y$ are reversed, as long as $n$ is bounded with the form
given by Eq. \ref{nmax}.
Moreover, for small $n''(\omega)$, we immediately see that the 
product of imaginary parts
$x''(\omega)y''(\omega)<0$
is negative. To be more specific, with the help of  Eq. \ref{one} we 
obtain that
\be\label{neg}
\eps''\mu''=\frac{1}{|z|^2}\left[(n''z')^2-(n'z'')^2\right]
\ee
One sees that the sign of $\eps''\mu''$ is fully determined by the 
r.h.s. of Eq. \ref{neg}.
We are not aware about any  physical requirement which prevents the 
r.h.s. of Eq. \ref{neg}
to be negative. In fact, data presented in Figs.2 and 3 show  clearly
that
\be\label{cond}
|n''z'|\le |n'z''|.
\ee
in the left part of the resonance gap, since  $|z'|<|z''|$ and $|n''|<|n'|$.
Thus, in agreement  with Eq. \ref{yy} we conclude that the opposite 
sign of $\epsilon''$
and $\mu''$ is a consequence of small transmission losses in the structure.
The same conclusion was derived in Ref. \cite{ms} for the DNG metamaterials.

\medskip

In conclusion, we presented the numerical analysis of the effective 
parameters of
two metamaterials: an array of SRR and an array of thin metallic cut wires.
We show that
the effective parameters of these systems exhibit resonant and 
anti-resonant behavior similar
to that found recently in the double negative metamaterials.
We suggest that the anti-resonant behavior is  caused by the requirement that
the refractive
index must be bounded in the structures which posses finite spatial 
periodicity.
As the spatial periodicity is an unavoidable property of the metamaterials, we
conclude that observed seemingly unphysical behavior of effective 
material parameters
is an intrinsic property of composites, which can not be avoided for instance
by decreasing the size of the unit cell.

\smallskip

The electromagnetic response of metamaterials is usually embodied in 
a description
involving bulk, continuous, frequency dependent permittivity and 
permeability tensors. This
description, however, is only approximate, as spatial dispersion is 
always present to
some degree in metamaterials.
Thus, the applicability of  our retrieval procedure, which uses  the 
formulas for transmission and
reflection of the homogeneous slab, might not be applicable in the neighbor
of the resonance frequency,  where  the wave length
of the electromagnetic wave inside the composite is already 
comparable with the spatial period.

Besides the analysis of the effective parameters, our results are 
interesting also for further
development of new double negative metamaterials, in which  the 
lattice of cut wires might find new
interesting applications
\cite{ms,Ekmel}.

\medskip

\noindent Acknowledgments.
This work was supported
by Ames Laboratory (Contract. n. W-7405-Eng-82). Financial support of DARPA
(Contract n. MDA972-01-2-0016),
NATO (Grant No. PST.CLG.978088),  VEGA (Project n.  2/3108/2003)
and EU$\underline{~~}$FET project DALHM are also acknowledged.

\begin{thebibliography}{99}
\bibitem{footnote1} Alternatively, we argue that the wavelength of 
the electromagnetic wave can not be smaller
than the spatial period of the structure, $\lambda>\lambda_{\rm min}=a$.
\end{thebibliography}
\end{document}